\documentclass[aps,prl,reprint,showpacs,groupedaddress]{revtex4-1}
\usepackage{amstext}
\usepackage{graphicx}
\usepackage{dcolumn}
\usepackage{bm}

\begin{document}


\title{Observation of Spin Flips with a Single Trapped Proton}

\author{S. Ulmer$^{1,2,3}$, C.C. Rodegheri$^{1,2}$, K. Blaum$^{1,3}$, H. Kracke$^{2,4}$, A. Mooser$^{2,4}$, W. Quint$^{3,5}$, J. Walz$^{2,4}$}
\affiliation{$^1$ Max-Planck-Institut f\"ur Kernphysik, Saupfercheckweg 1, D-69117 Heidelberg, Germany}
\affiliation{$^2$ Institut f\"ur Physik, Johannes Gutenberg-Universit\"at D-55099 Mainz, Germany}
\affiliation{$^3$ Ruprecht Karls-Universit\"at Heidelberg, D-69047 Heidelberg, Germany}
\affiliation{$^4$ Helmholtz Institut Mainz,  D-55099 Mainz, Germany}
\affiliation{$^5$ GSI - Helmholtzzentrum f\"ur Schwerionenforschung, D-64291 Darmstadt, Germany}

\date{\today}

\begin{abstract}
Radio-frequency induced spin transitions of one individual proton are observed for the first time. The spin quantum jumps are detected via the continuous Stern-Gerlach effect, which is used in an experiment with a single proton stored in a cryogenic Penning trap. This is an important milestone towards a direct high-precision measurement of the magnetic moment of the proton and a new test of the matter-antimatter symmetry in the baryon sector.
\end{abstract}


\maketitle

\begin{figure}[htb]
        \centerline{\includegraphics[width=6.5cm,keepaspectratio]{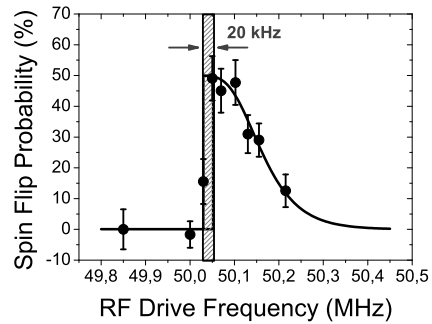}}
            \caption[FEP]{Proton spin flip resonance in the analysis trap with its inhomogeneous magnetic field.}
            \label{fig:SFProb}
\end{figure}

The challenge to understand the structure of the proton and to measure its properties inspires very different branches of physics. A very important property of the particle is its magnetic moment $\mu_\text{p} = (g_\text{p} / 2) \mu_N$, where $g_\text{p}$ is the Land\'e \emph{g}-factor, $\mu_\text{N} = e \hbar / (2 m_\text{p})$ is the nuclear magneton, and $e/m_\text{p}$ is the charge-to-mass ratio. The most precise value for the \emph{g}-factor of the free proton $g_\text{p}=5.585\,694 \,706(56)$  is extracted theoretically from a measurement of the magnetic field dependence of the hyperfine splitting in hydrogen by means of a maser \cite{winkler1972magnetic,KB}. The precision achieved by that experiment is 1 part in $10^8$, limited by the ``wall shift'' due to the interaction of the hydrogen atoms with the maser cavity.\\
Our experiment aims at a direct measurement of the magnetic moment of one individual proton stored in a Penning trap \cite{CCR, Klaus}. The particle is confined to a volume below 100$\,\mu$m$^3$ by means of electrostatic and magnetostatic fields. Thus, the experiment is exceptionally ``clean'' in the sense that interactions with the apparatus can be neglected, no electron is involved, and thus, no theoretical corrections are required. It should be possible to improve the accuracy of the proton \emph{g}-factor by one order of magnitude, at least.\\
This opens up a highly exciting perspective; the same techniques can also be used to measure the magnetic moment of the antiproton \cite{quint1993mma}, which is presently known at a level of $10^{-3}$, only \cite{Pask}.  Thus, an improvement by six orders of magnitude is possible, which will represent a very stringent test of the matter-antimatter symmetry (\emph{CPT}-symmetry) in the baryon sector \cite{Bluhm}.\\
The \emph{g}-factor determination reduces to the measurement of two frequencies, $g_\text{p}=2\nu_\text{L}/\nu_\text{c}$. Both frequencies, the free cyclotron frequency $\nu_c=e B/(2\pi m_\text{p})$, and the Larmor frequency $\nu_L=g_\text{p} e B/(4\pi m_\text{p})$, are defined by the charge-to-mass ratio of the proton, and the magnetic field $B$ of the Penning trap.  The free cyclotron frequency is obtained via the relation $\nu_\text{c}^2={\nu_+^2+\nu_-^2+\nu_z^2}$, where $\nu_+, \nu_-$ and $\nu_z$ are the respective eigenfrequencies of the single proton stored in the Penning trap \cite{brown1986gtp}. The measurement of the Larmor frequency $\nu_\text{L}$ is based on the continuous Stern-Gerlach effect \cite{dehmelt}. This elegant scheme for the non-destructive determination of the spin direction has been used with great success in measurements of the \emph{g}-factor of the electron \cite{Hanneke}, the positron \cite{vandyckjr1987nhp} and of bound electrons \cite{Hermanspahn, Verdu}.  All these experiments dealt with magnetic moments on the level of the {\em{Bohr magneton}}.  A measurement of the magnetic moment of the proton is much more challenging, because it is on the level of the {\em{nuclear magneton}}, which is about three orders of magnitude smaller.  \\
In this Letter we demonstrate the detection of spin flips of a single proton stored in a Penning trap, which is the most important milestone for our experiment.\\
A double Penning trap \cite{CCR}
\begin{figure}[htb]
        \centerline{\includegraphics[width=9cm,keepaspectratio]{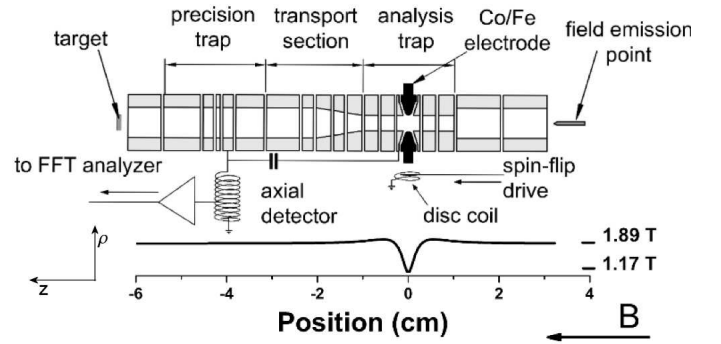}}
            \caption[FEP]{Schematic of the proton \emph{g}-factor experiment. The system consists of two Penning traps which are connected by transport electrodes. The central ring electrode of the analysis trap is made of ferromagnetic material. The lower graph shows the magnitude of the magnetic field along the \emph{z}-axis. For further details see text.}
            \label{fig:TrapDetect}
\end{figure}
consisting of a precision trap (PT) and an analysis trap (AT) is mounted in the horizontal bore of a superconducting magnet with a magnetic field of $1.89\,$T. Both traps are of cylindrical and compensated design \cite{gabrielse1989oep}. The precision trap has an inner radius of $3.5\,$mm and is located in the homogeneous center of the magnetic field. The ring electrode of the analysis trap has an inner radius of $1.8\,$mm and is made out of ferromagnetic Co/Fe-material which distorts the homogeneous magnetic field. Near the center of the trap the magnetic field is approximately described by
    \begin{eqnarray}
    \vec{B}=B_0 \vec{e}_z+B_2\left((z^2-\rho^2/2)\vec{e}_z- z\rho\vec{e}_\rho\right)~.
    \label{eq:Bottle}
    \end{eqnarray}
The ``magnetic bottle'' coefficient $B_2=3.00(10)\cdot10^5\,$T$\cdot$m$^{-2}$ has been determined by shifting a single proton through the trap and measuring its cyclotron frequency. $B_2$  is $3.8$ times larger than that used in a competing experiment \cite{Nick} and is the highest magnetic field inhomogeneity ever superimposed to a Penning trap. \\
The two traps are connected by transport electrodes. The inner radius reduces along these electrodes to match the different trap radii, as shown in Fig.\ \ref{fig:TrapDetect}. Voltage ramps are applied to the electrodes of the transport section to move the proton adiabatically between both Penning traps. The whole electrode stack is mounted in a sealed vacuum chamber and cooled to $3.8\,$K with a pulse tube cooler. Due to cryo-pumping the background pressure in such systems is $<10^{-14}\,$Pa. We can store a proton in our trap for arbitrarily long times (months).  \\
Protons are produced in the precision trap by electron bombardment of a polyethylene-target and subsequent electron impact ionization. Impurity ions are selectively heated out of the trap using broadband noise and resonant radio frequency (rf) drives. Particles are then removed from the clean proton cloud until just one single particle remains. The trapped proton oscillates at three different eigenmotions \cite{brown1986gtp}, the modified cyclotron motion at $\nu_\text{+,PT}\approx28.97\,$MHz, the axial motion at $\nu_\text{z}\approx674\,$kHz, and the magnetron motion at $\nu_\text{-,PT}\approx 8\,$kHz. The oscillation of the particle induces image currents in the trap electrodes which are detected as rf voltages across a resistance $R$. In practice a coil is used instead of a resistor. The inductance together with the stray capacitance of the trap system forms a parallel tuned circuit with a quality factor \emph{Q}. The axial frequency of the stored proton in the precision trap and in the analysis trap is measured with a superconducting Nb/Ti coil connected to both traps. The quality factor is about $5700$ and the effective on-resonance resistance is $36\,$M$\Omega$ \cite{ulmer2009quality}. The detection system for the measurement of the modified cyclotron frequency in the precision trap (not shown in Fig. \ref{fig:TrapDetect}) is made out of copper. It has a quality factor of $1250$ and an on-resonance resistance of $380\,$k$\Omega$. The voltage signals induced by the proton are amplified with cryogenic low-noise amplifiers and analyzed with an FFT spectrum analyzer. The magnetron frequency is measured by sideband coupling \cite{Cornell} to the axial eigenmotion. The proton interacts with the tank circuits and thus cools resistively to ambient cryogenic temperatures.
The magnetic field in the center of the analysis trap is $1.17\,$T due to the field distortion by the ferromagnetic ring electrode. The cyclotron frequency of the proton is measured as described in \cite{Hermanspahn} and is $\nu_\text{+,AT}\approx17.91\,$MHz. The magnetron frequency in the analysis trap is $\nu_\text{-,AT}\approx13\,$kHz. \\
In an ideal Penning trap the axial frequency $\nu_z$ of a trapped proton is
    \begin{eqnarray}
    \nu_{z,0}=\frac{1}{2\pi}\sqrt{\frac{e}{m_p}\frac{d}{dz}\left(-\vec\nabla\Phi_\text{E}|\vec{e}_z\right)}=\frac{1}{2\pi}\sqrt{\frac{e V_0}{m_pd^2}}~,
    \end{eqnarray}
where $d$ is a characteristic length of the trap, $V_0$ is the trapping voltage and $e\Phi_\text{E}$ is the electrostatic energy. In the analysis trap with its strong magnetic bottle the magnetostatic energy $\Phi_\text{M}=- \vec\mu \cdot \vec B$ contributes to the axial frequency.
Here $\vec \mu$ is the sum of the Landau-magnetic moment, which is due to the angular momenta of the cyclotron and the magnetron motion, respectively, and the magnetic moment of the proton spin $\mu_s=(g_\text{p}e\hbar m_s)/(2m_\text{p})$, where  $m_s$ is the spin quantum number. The energy $\Phi_\text{M}$ contains a term proportional to $z^2$ where $z$ is the axial distance to the center of the trap. The axial frequency thus becomes $\nu_\text{z}=\nu_\text{z,0}+\Delta\nu_\text{z}(n_+,n_-,m_s)$, where $n_\pm$ denotes the number of quanta in the radial modes, and
    \begin{eqnarray}
    \Delta\nu_z(n_+,n_-,m_s)&=&\nonumber\\
    \frac{h \nu_+}{4\pi^2 m_\text{p}\nu_z}\frac{B_2}{B_0}&\cdot&\left(n_++\frac{1}{2}+\frac{\nu_-}{\nu_+}\left(n_-+\frac{1}{2}\right)+\frac{g_\text{p}m_s}{2}\right)~.
    \label{eq:Bottleshift}
    \end{eqnarray}
A cyclotron quantum jump $\Delta n_+=\pm 1$ corresponds to a radial energy change of $\Delta E_+=\pm74\,$neV and causes an axial frequency shift of $\Delta\nu_z=\pm68\,$mHz. A transition of the magnetron quantum number $\Delta n_-=\pm 1$ leads to $\Delta\nu_z=\pm49\,\mu$Hz. Changes of both radial quantum numbers $\Delta n_\pm=\pm 1$ are due to electric dipole transitions. A magnetic-dipole spin-flip transition $\Delta m_s = \pm 1$ causes a jump of the axial frequency $\Delta\nu_\text{z,SF}=\pm190\,$mHz. This is most important because the axial frequency can thus be used to detect the spin direction of the proton.\\
Spin-flip transitions are driven using a disc coil mounted close to the electrode stack as shown in Fig.\ \ref{fig:TrapDetect}.
The coil generates a transverse magnetic rf-field $\vec{b}_\text{rf}$ with frequency $\nu_\text{rf}$. To drive spin flips $\vec{b}_\text{rf}$ is tuned near to the Larmor frequency $\nu_\text{L}=g_\text{p}\nu_c/2$. The field penetrates into the trap through slits between the electrodes. This rf field causes a precession of the proton spin around the $\vec{b}_\text{rf}$-axis at the Rabi frequency $\Omega_R/2\pi=\nu_\text{L} b_\text{rf}/B_0$. In the presence of a strong magnetic bottle $B_2$ the spin-flip probability is given by \cite{Brown2}
    \begin{eqnarray}
    P_\text{SF}=\frac{1}{2}\left(1-\exp\left(-\frac{1}{2}\Omega_R^2 t_0~\chi(2\pi\nu_\text{rf},B_2,T_z)\right)\right)~,
    \label{eq:SFProb}
    \end{eqnarray}
where $t_0$ is the irradiation time, $\chi(2\pi\nu_\text{rf},B_2,T_z)$ the transition lineshape and $T_z$ the axial temperature.\\
The axial frequency $\nu_z$ has a strong dependence on the radial energy $E_\rho=|E_+|+|E_-|$ with $(\Delta\nu_z)/(\Delta E_\rho) \approx0.92\,$Hz/$\mu$eV, where $|E_\pm|$ is the energy of the cyclotron mode and the magnetron mode, respectively. It is thus essential to avoid induced spurious changes of the radial energy of the proton. Care is taken to shield the apparatus from perturbing rf-signals and noise. In the lines for the spin-flip drive a chain of band-pass filters attenuates spurious rf-fields at the cyclotron frequency by $100\,$dB. The line is shorted to ground with an rf-relay whenever possible.\\
\begin{figure}[htb]
        \centerline{\includegraphics[width=6.5cm,keepaspectratio]{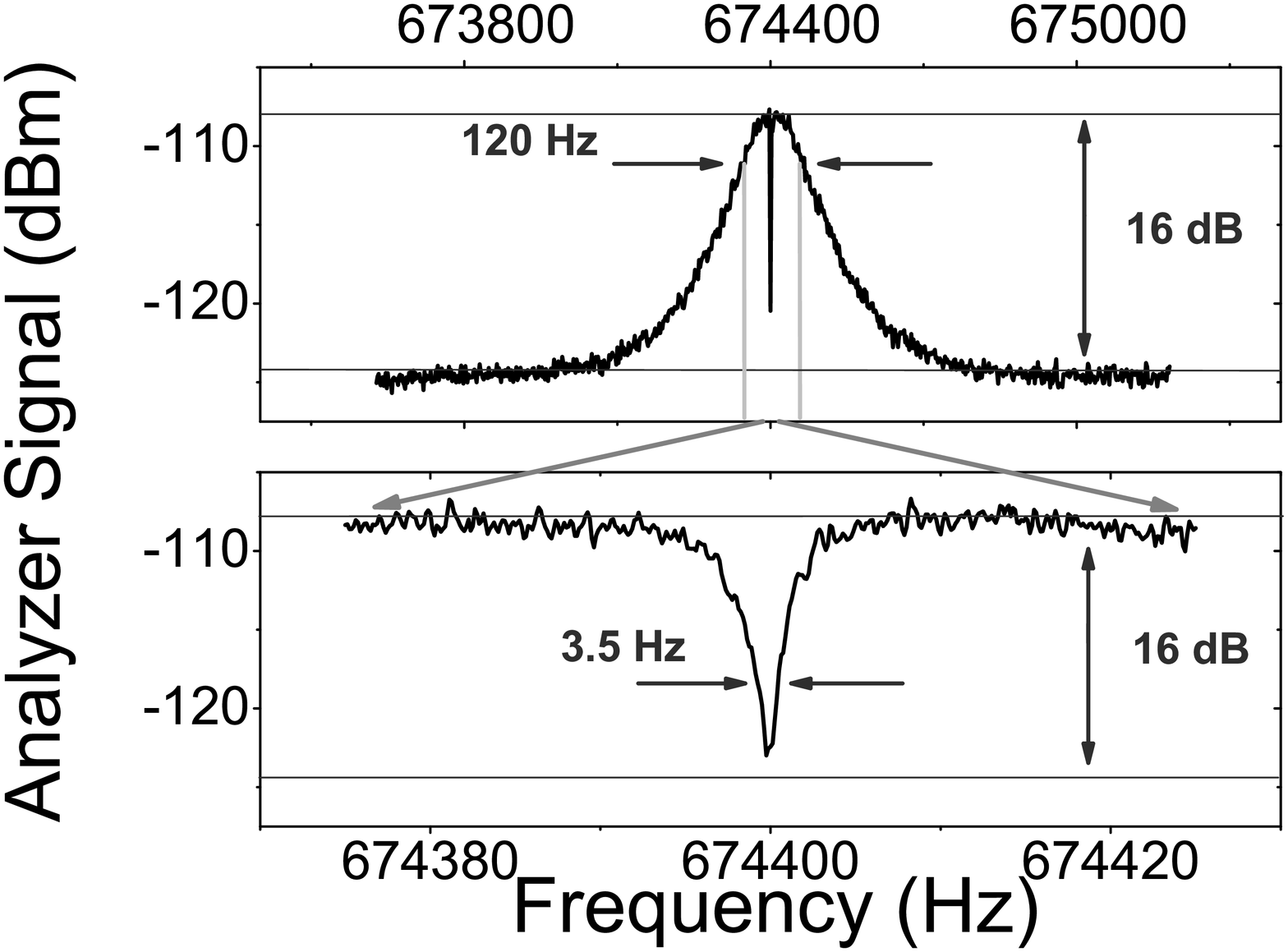}}
            \caption[FEP]{Spectrum of the axial detector. The peak in the upper graph is the thermal noise spectrum of the cryogenic tank circuit. Its width is $\nu_z/Q\approx 120\,$Hz. The frequency axis is stretched in the lower graph to show the narrow dip, which is due to the trapped proton shorting the noise of the detector.}
            \label{fig:CauseRes}
\end{figure}
The frequency $\nu_z$ is measured by observing the thermal noise spectrum of the cryogenic axial tank circuit, see Fig.\ \ref{fig:CauseRes}. The particle acts like a series tuned circuit, shorting the thermal voltage-noise $e_\text{th}=\sqrt{4k_\text{B} T_z R\Delta F}$ of the detector, where $\Delta F$ is the FFT spectrum analyzer bandwidth. The linewidth of the dip is $1/(2\pi\tau_z)$, where $\tau_z=(m_\text{p}/R)\cdot(D^2/q^2)$ is the cooling time constant and $D$ is a characteristic length of the trap. In the precision trap the signal linewidth is $1.25\,$Hz and in the analysis trap $3.5\,$Hz due to the smaller trap-size. The detection system has a signal-to-noise ratio of 16$\,$dB.\\
To determine the spin direction of the proton, the axial frequency $\nu_z$ has to be measured with high resolution.
The axial frequency resolution achieved by noise-dip detection increases with measuring time $\Delta t$ due to spectrum analyzer averaging. On the other hand, tiny perturbing effects can drive the radial modes and cause small changes of the axial frequency which increase with time. It was found in a series of experiments that fluctuations of the cyclotron energy at an average rate of 0.045 quantum jumps per second are responsible for the residual instability of the axial motion. Details of these experiments will be discussed in a forthcoming publication.\\
In the following discussion the change of the axial frequency between time $t$ and $t+\Delta t$ is considered and the standard deviation of $\alpha=\nu_z(t)-\nu_z(t+\Delta t)$ is considered. We quantify the frequency fluctuation by using the standard deviation of $\alpha$, namely $\Xi=\left((N-1)^{-1}\sum(\alpha_i-\bar\alpha)^2\right)^{1/2}$. In our experiment at an FFT averaging time of $\Delta t_\text{opt}=80\,$s the frequency fluctuation is minimal, $\Xi_\text{min}\approx150\,$mHz. This value is not yet sufficiently stable to detect spin flips directly in one measurement sequence as described in \cite{Hartmut}. However, averaging can be used to detect proton spin flips.\\
For such a statistical measurement the axial frequency $\nu_z$ is
determined in sequences of three measurements $\nu_{z,1}$,
$\nu_{z,2}$, and $\nu_{z,3}$. Between the first and the second
measurement a spin-flip drive is turned on near the Larmor frequency
$\nu_\text{L}\approx\nu_\text{rf}=50.102\,$MHz. Between the second
and the third measurement the rf-synthesizer is tuned to a reference
frequency 100$\,$kHz below $\nu_\text{L}$ which corresponds to
approximately one linewidth. Between the third measurement and the
first measurement of the following cycle $\nu_{z,1'}$ no rf-signal
is applied to the trap. This sequence is repeated several hundred
times and the standard deviations of the frequency differences $\Xi_\text{SF}=\nu_{z,2}-\nu_{z,1}$,
$\Xi_\text{ref}=\nu_{z,3}-\nu_{z,2}$ and
$\Xi_\text{back}=\nu_{z,1'}-\nu_{z,3}$ are computed. If spin flips
are driven the corresponding axial frequency shifts add up in a
statistical way. If the spin of the proton flips in \emph{M} cycles
out of \emph{N}, the total fluctuation is
    \begin{eqnarray}
    \Xi_\text{SF}&=&\sqrt{\sum_i^M\frac{\left(\alpha_i\pm\Delta\nu_{z,SF}-\bar\alpha\right)^2}{N-1}
    +\sum_{i=M}^N\frac{\left(\alpha_i-\bar\alpha\right)^2}{N-1}}\nonumber\\
    &\approx&\sqrt{\Xi_{back}^2+P_{SF}\Delta\nu_{z,SF}^2}~,
    \label{eq:SFBack}
    \end{eqnarray}
where $P_\text{SF}$ is given by Eq.\ (\ref{eq:SFProb}).
The comparison of $\Xi_\text{ref}$ and $\Xi_\text{back}$ is a means to test for spurious heating of the cyclotron motion due to rf-synthesizer noise, which might increase the axial frequency fluctuation. If both quantities are the same within error bars this indicates that the cyclotron motion is not affected and that an increase of $\Xi$ is due to proton spin flips. A resolution limit for the detection of spin flips by this method is reached if the error $\sigma_{\Delta\Xi}=((\Xi_\text{ref}/\sqrt{2N-2})^2+(\Xi_\text{SF}/\sqrt{2N-2})^2)^{0.5}$ is of the same size as the increase of the frequency fluctuation $\Delta\Xi=\Xi_\text{SF}-\Xi_\text{ref}$. With the conditions of our experiment for $P_\text{SF}=50\,\%$, $\Delta\Xi\approx50\,$mHz are expected. This shows that it is possible to statistically detect spin flips in a series of only fifty measurement cycles.\\
\begin{figure}[htb]
        \centerline{\includegraphics[width=8.0cm,keepaspectratio]{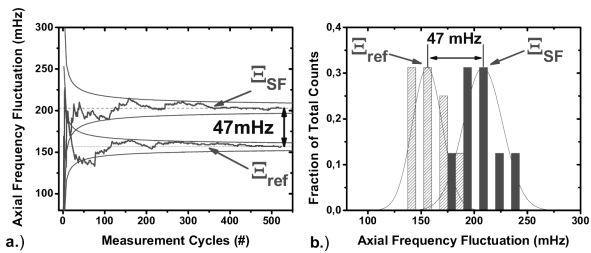}}
            \caption[FEP]{Proton spin flip data. In a.) the evolution of the frequency fluctuations $\Xi_\text{SF}$ and $\Xi_\text{ref}$ as a
    function of measurement cycles is shown. The clear separation of the
    two lines shows that spin flips are detected. An alternative
    analysis is shown in b.). The entire data sequence was split into
    subsets and the values of $\Xi_\text{SF}$ and $\Xi_\text{ref}$ were
    binned to two histograms.}
            \label{fig:Result}
\end{figure}
In the experiment the measurement sequence is repeated several hundred times and the frequency fluctuations $\Xi_\text{SF}$ and $\Xi_\text{ref}$ are determined. The evolution of both quantities as a function of measurement cycles is shown in Fig.\ \ref{fig:Result} a.) along with a confidence band given by $\sigma_{\Delta\Xi}$. Both fluctuations converge for large measuring times to nearly constant values. A significant shift of $\Delta\Xi=47\,$mHz between the reference measurement $\Xi_\text{ref}$ and the resonant measurement $\Xi_\text{SF}$ is observed. This means that spin flips are detected in the experiment.
Using Eq.\ (\ref{eq:SFBack}) we find an approximately saturated spin-flip probability of $P_\text{SF}=47\pm7\,\%$. The data set has also been analyzed in a different way. The whole sequence of 800 data points is split into subsequences of 50 measurements. For every subsequence $\Xi_\text{SF}$ and $\Xi_\text{ref}$ are determined and binned into histograms. Figure \ref{fig:Result} b.) shows a distinct splitting between both distributions. This clearly shows, again, that proton spin flips are detected.\\
A spin flip resonance of the proton in the analysis trap has been measured by tuning $\nu_\text{rf}$ in steps across the Larmor frequency $\nu_\text{L}$. For every frequency point the data are analyzed as above and the corresponding spin-flip probability $P_\text{SF}$ is determined. In the magnetic bottle the Larmor frequency $\nu_\text{L}$ is a function of the particle position $z$. Due to the axial oscillation $z(t)$ the spin-flip resonance curve has a broad width. It is asymmetric due to the Boltzmann distribution of the axial energy. Figure \ref{fig:SFProb} presents the spin-flip resonance. The solid line is the best fit of Eq.\ (\ref{eq:SFProb}) to the data. Fixed parameters are the irradiation time $t_0=10\,$s, the magnetic bottle strength $B_2=3.00(10)\cdot10^5\,$T$\cdot$m$^{-2}$, and the axial temperature of $T_z=9.5\,$K. $B_2$ and $T_z$ have been measured independently. Free fit parameters are the amplitude $b_\text{rf}$ of the magnetic rf-field, which comes out as $2.5\,\mu$T, and the Larmor frequency $\nu_\text{L}$. From the fit the Larmor frequency can be determined with a precision of 2$\cdot$10$^{-4}$. The result is consistent with the independent measurement of the magnetic field in the analysis trap using the proton's cyclotron frequency. The agreement between the data and the fit for the variation of the spin-flip probability with the drive frequency provides convincing evidence that proton spin flips were detected. \\
In the future we plan to drive the spin flip in the precision trap and to transport the particle into the analysis trap to analyze its spin direction \cite{Hartmut}. In the precision trap the magnetic field is four orders of magnitude more homogeneous than in the analysis trap and a narrow spin-flip resonance is expected. Together with a measurement of the cyclotron frequency this will be a direct high-precision measurement of the \emph{g}-factor of the proton. \\
In conclusion, spin flips of a single trapped proton have been observed for the first time.  A strong magnetic bottle $B_2$ has been used to couple the proton spin to its axial motion in a Penning trap.  The observation of tiny changes of the axial frequency clearly indicates  proton spin flips.  A Larmor resonance was clearly observed by averaging.
This is an important milestone towards a direct high-precision measurement of the magnetic moment of the proton and the antiproton.
\\
We acknowledge the support of the Max-Planck Society, the BMBF, the DFG (QU-122-3), the Helmholtz-Gemeinschaft, HGS-HIRE, Al$\beta$an (E06D101305BR) and the IMPRS-QD.

\end{document}